# Health diagnosis and recuperation of aged Li-ion batteries with data analytics and equivalent circuit modeling

Riko I Made[a], Jing Lin[b], Jintao Zhang[a], Yu Zhang[b], Lionel C. H. Moh[a], Zhaolin Liu[a], Ning Ding[a], Sing Yang Chiam[a], Edwin Khoo[b,*], Xuesong Yin[a,*] and Guangyuan Wesley Zheng[c]*

[a] Institute of Materials Research and Engineering (IMRE), Agency for Science, Technology and Research (A*STAR), 2 Fusionopolis Way, Innovis #08-03, Singapore 138634, Republic of Singapore.
Email: yinxs@imre.a-star.edu.sg
[b] Institute for Infocomm Research (I²R), Agency for Science, Technology and Research (A*STAR), 1 Fusionopolis Way, #21-01 Connexis, Singapore 138632, Republic of Singapore.
Email: edwin_khoo@i2r.a-star.edu.sg
[c] Posh Robotics, 3501 Breakwater Court, Hayward, CA 94545, United States.
Email: wesley@poshrobotics.com

**Abstract**

Battery health assessment and recuperation play a crucial role in the utilization of second-life Li-ion batteries. However, due to ambiguous aging mechanisms and lack of correlations between the recovery effects and operational states, it is challenging to accurately estimate battery health and devise a clear strategy for cell rejuvenation. This paper presents aging and reconditioning experiments of 62 commercial high-energy type lithium iron phosphate (LFP) cells, which supplement existing datasets of high-power LFP cells. The relatively large-scale data allow us to use machine learning models to predict cycle life and identify important indicators of recoverable capacity. Considering cell-to-cell inconsistencies, an average test error of 16.84% ± 1.87% (mean absolute percentage error) for cycle life prediction is achieved by gradient boosting regressor given information from the first 80 cycles. In addition, it is found that some of the recoverable lost capacity is attributed to the lateral lithium non-uniformity within the electrodes. An equivalent circuit model is built and experimentally validated to demonstrate how such non-uniformity can be accumulated, and how it can give rise to recoverable capacity loss. SHapley Additive exPlanations (SHAP) analysis also reveals that battery operation history significantly affects the capacity recovery.

**Introduction**

Large numbers of Li-ion batteries (LIBs) are produced to meet the ever-growing demands of the electric vehicle (EV) market up to an annual capacity of 2500 GWh by 2030.[1] These EV batteries will eventually retire and accumulate after their service due to a limited cycle life.[2,3] The precious and hazardous nature of LIB components necessitates materials recycling to attain a sustainable battery industry and circular economy.[4-7] Meanwhile, considerable residual capacities after EV applications, normally above 70% of the initial capacities, bring about an emerging field of repurposing and remanufacturing of retired EV batteries for less demanding applications, like low-speed vehicles or energy storage systems.[8,9] Therefore, providing a second life to retired batteries becomes an economically viable method to extend their service life before materials recycling (Fig. 1).

State of health (SoH), which is defined by the ratio of residual to initial capacity ($\text{SoH} = C_{res}/C_{ini}$) and describes the aging status, is an essential parameter to be understood before any repurposing activities.[10,11] However, the ambiguous aging mechanisms of Li-ion batteries make it challenging to estimate SoH accurately. Data-driven methodologies show great potential in dealing with complex systems with non-linear behaviors. Several works have been carried out to accurately estimate cycle life of Li-ion batteries by machine learning (ML) methods.[12-16] Cell chemistries and operational conditions, such as charge/discharge currents, cut-off voltages, depth of discharge and temperature, have been considered in generating datasets. Nevertheless, the intrinsic inconsistencies in cell performance that result from the imperfection of cell manufacturing are often overlooked by assuming the cells of the same batch or manufacturer are identical. In practice, even batteries from the same batch will have different SoHs when used in a battery pack, because individual cells in a pack rarely work under an ideally equivalent operational or environmental condition.[17] For second-life batteries, this inconsistency problem becomes more significant when various cells with different operational histories need to be re-grouped and employed in a new system. Therefore, cell consistency is an important factor to examine when assessing the SoH of Li-ion batteries.

In addition to an accurate estimation of the SoH, it is also beneficial to recuperate some lost capacity of the retired cells before a second-life application. This task is also highly dependent on the aging mechanisms. Various and mixed degradation modes have been proposed to understand the aging behaviors of Li-ion batteries,[11,18-20] but only those involving reversible changes may contribute to a capacity recovery.

Electrolyte consumption has been considered as an aging factor, and re-filling electrolyte was reported to achieve a capacity recovery over 10%.[21] Direct lithiation from an external lithium source was also proven effective to compensate lost Li inventory in the cathode and recover some capacity.[22] However, the invasive nature of these methods posts challenges towards a large-scale and reliable process that can be readily adopted by industry. For non-invasive methods, thermal treatment was proposed to break and re-build the solid-electrolyte interface (SEI) layer, whose continuous growth during cycling was believed to be the reason for loss of active Li and increase of cell impedance.[23] To partially reverse degradation attributed to a non-uniform distribution of Li ions, extreme voltage reconditioning was carried out to re-distribute and activate the inactive Li ions.[24] Although discrete trials on individual cells were reported, there lacks a comprehensive evaluation of the effectiveness of cell capacity recovery techniques with a sufficient sampling quantity where the cell inconsistency is considered.

Lithium iron phosphate (LFP) batteries are a popular candidate in various energy storage applications because of its long cycle life, high safety and low cost.[25] The relatively low value of the raw materials makes recycling of LFP batteries less profitable than that of lithium nickel manganese cobalt oxide (NMC) cells.[26] From this aspect, it is essential that second-life applications can make better use of retired LFP cells before they are sent to recycling.[27]

In this work, we collected cycling data of 62 LFP 18650 cells cycled under various operational conditions, such as current, cycle number, etc. Data analytics and Gaussian process modeling were used to predict the cell cycle life given the cycling condition and SoH of early cycles. The non-uniformity in battery cells, such as uneven distributions of active materials, variations of electrode thickness and other manufacturing imperfections, was hypothesized as a reasonable cause of recoverable performance degradation. An equivalent circuit model (ECM) was employed to quantitatively explore the mechanisms of capacity loss and the feasibility of capacity recovery. The experimental reconditioning treatments applied to the aged cells validated their effectiveness with some of the lost capacity recovered. Finally, the importance of different operational parameters to the cell capacity recoverability was assessed by SHapley Additive exPlanations (SHAP) analysis[28], which suggested the significant role of SoH, residual capacity and the cycling current. In summary, we have made three main contributions in this work. First, we have performed large-scale LFP cell aging and reconditioning experiments that enable the use of ML models to predict cycle life. Second, we use an ECM, which is validated using the experimental data collected, to show how lateral lithium non-uniformity in electrodes can be built up due to an uneven internal resistance, and how such non-uniformity causes capacity loss that can be recovered by extreme voltage holds. Third, we identify important parameters that affect the magnitude of capacity recuperation, where reconditioning eliminates some of the Li imbalance.

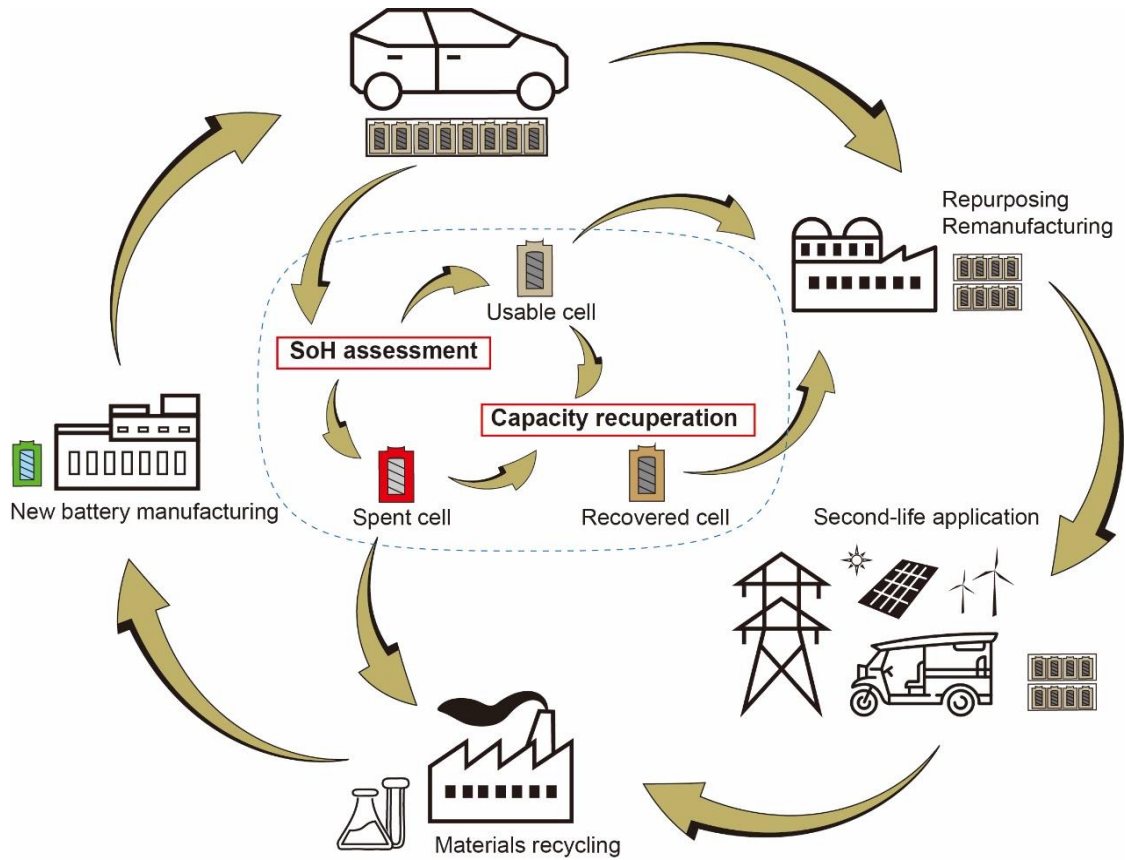

Fig. 1. Circular utilization of Li-ion batteries with emphasis on state of health (SoH) assessment and capacity recuperation for second-life applications.

**Experimental**

Battery cycling

Commercially available LFP/graphite 18650 cells with a nominal capacity of ~1.5 Ah (high-energy type) were used in this study. This type of cells has a higher capacity but shorter cycle life in contrast to the LFP 18650 cells used in a previous study (high-power type) by Severson et al.[12] Even for the same LFP chemistry, more comprehensive data from diverse cells are helpful for developing robust data-driven methodologies and understanding cell inconsistencies. Constant current charge and constant current discharge cycling with a voltage window from 2.5 V to 4.0 V was applied to perform cell aging. The charge-discharge current was varied at 1.0 A, 1.5 A, 2.0 A and 3.0 A to investigate its effects on cell aging and recovery. The testing was carried out on a 16-channel battery tester (Maccor, Model-4200) in ambient environment. The photos of battery cells and testing setup are presented in Supplementary Information Fig. S1. The details of cell cycling information are listed in Supplementary Information Table S1.

Capacity recuperation

The accumulated non-uniform distribution of lithium ions upon cycling is believed to be one of the causes of the capacity fade. Applying an electric voltage for an extended period of time would provide a driving force to re-homogenize the lithium ions in the electrodes and recover some of the lost capacity. In this regard, the aged cells with various SoHs (40-80%) were subjected to constant voltage reconditioning treatments. The cells were kept at a charged state (3.6 V) and followed by a discharged state (2.0 V) for 72 h each. This procedure was applied to all the cells with different aging histories. The cell capacities before and after the treatments were checked by cycles of a constant current charge and a constant current discharge at a small current of 0.3 A. Typical voltage and current curves of the recuperation and capacity check processes are shown in the Supplementary Information Fig. S2. The recovery rates were calculated as the ratio of recovered capacity to both the residual and lost capacity.

Data-driven assessment of battery states

The discharge capacity data were batch processed to build the distributions of the cycle life. We set the discharge capacity of 1.1 Ah as the cutoff value at end of first life to define the cycle life -- for an acceptable SoH around 73% (1.1 Ah/1.5 Ah) for the cells (Fig. 2a). We consider several combinations of features to build a model to predict cycle life, which are the number of cycles to reach a discharge capacity of 1.1 Ah, the cycling current, two early discharge capacities at 'start' cycle number $n_{start}$ and 'end' cycle number $n_{end}$, and variance of discharge curve difference between two cycles $var(\Delta Q(U))$.[12]

We used linear regressor (LR), Gaussian process regressor (GPR) and gradient boosting regressor (GBR)[29] to predict the cycle life over our 62-cell dataset. Although a sample size of 62 cells is relatively large in the academic battery community, it is relatively small for ML model fitting and therefore, our models are prone to overfitting. To reduce overfitting in commonly observed in data-driven models trained on small datasets, we performed nested cross-validation (CV), which consists of an outer loop for evaluating test errors and an inner loop for hyperparameter tuning as shown in Supplementary Information Fig. S3. The outer CV loop splits the data into 5 sets of train-validation and test sets, with an index running from 0 to 4 (Supplementary Information Fig. S3). The inner CV loop further splits each train-validation set into 5 sets of train and validation sets, which are used for hyperparameter tuning.

For hyperparameter tuning, a grid search was conducted within the inner loop over the ($n_{start}$, $n_{end}$) hyperparameters to reduce the mean square error (MSE) on the validation set. An optimal set of ($n_{start}$, $n_{end}$) from this inner loop hyperparameter tuning is selected such that it gives the lowest average MSE on the 5 inner loop validation sets. Using this optimal set of ($n_{start}$, $n_{end}$), the model is finally retrained on the train-validation set. At the end of the entire nested CV campaign, we had 5 tuned models that correspond to the 5 train-validation/test sets in the outer loop. The final model performance is quantified by the average of the 5 root mean square errors (RMSEs) and the average of 5 mean absolute percentage errors (MAPEs) evaluated on the train-validation and test sets. These test error metrics provide an estimate of the generalization performance of the model in predicting cycle life.

Simulation of capacity loss and recovery using equivalent circuit modeling

Lateral electrode inhomogeneity of lithium-ion cells has been widely studied in literature to better understand its causes and characteristics.[24] Such inhomogeneity has been attributed to manufacturing variations and defects, certain cell form factors, and tab locations, which in turn give rise to a non-uniform current density and hence uneven lithium ion distribution. Some efforts have also focused on simulating the formation of such a inhomogeneity using physics-based and circuit-based models[30], by dividing a cell laterally into multiple local mini-cells and accounting for the internal resistance associated with the interconnection among these mini-cells. However, there is little work that studies how such inhomogeneity affects the usable capacity, and how capacity loss due to inhomogeneity may be recovered by homogenizing the cell using certain operations.

Spingler and coworkers[24] proposed that shallow cycling around the mid state-of-charge (SOC) range with shallow electrode open circuit voltage (OCV) curves can significantly encourage inhomogeneity to develop and reduce usable capacity within a voltage window. They also showed experimentally that capacity loss can be recovered by holding the cell for several days at extreme voltage at which the electrode OCV curves are much steeper. They proposed that such a phenomenon might be captured by an ECM but have not implemented and tested this idea. In this work, we implement a simple ECM to simulate the recoverable capacity loss due to lithium inhomogeneity and its recovery under extreme voltage reconditioning, and we compare the simulation results to experimental measurements.

To demonstrate the formation of lithium inhomogeneity and the effects of electric reconditioning on the capacity recovery of aged LFP batteries, an ECM simulation is carried out. We divide the cell laterally into two parallel sub-cells of equal capacity. For each sub-cell, we model the anode and cathode open circuit potential separately ($U_{OCV,1}$, $U_{OCV,2}$). This contrasts with the monolithic full-cell OCV used by most ECMs in literature. Moreover, we associate an internal resistance ($R_1$, $R_2$) with each electrode and add a bridge connecting the two sub-cells that accounts for the electrolyte resistance ($R_e$), which lithium ions need to overcome to transport laterally across different sub-cells. This bridge $R_e$ is what enables the lithium distribution to become uneven when one sub-cell somehow has lower resistance ($R_2^\pm < R_1^\pm$), so lithium ions may prefer to be cycled within that branch, driving the lateral transport through the bridge. Since the electrodes are manufactured to be thin but with a large cross-section to minimize internal resistance while maintaining substantial capacity, the lateral resistance $R_e$ at the bridge is typically much larger than the thickness-wise resistances ($R_1^\pm$ and $R_2^\pm$). This ECM is the most parsimonious model that can simulate the effects of lateral lithium inhomogeneity.

To find suitable values for the ECM parameters ($R_1$, $R_2$ and $R_e$), we fit the ECM to the experimental cell terminal voltage curves for the last few cycles and the whole recuperation process, including capacity checkup steps. We solve this nonlinear least squares problem using the differential evolution algorithm as implemented in the Python library SciPy.[31]

Capacity recovery statistics

In this section we are interested in quantifying the recoverable capacity, as well as the factors that contribute to the cell's capacity recoverability. We define several metrics to quantify the recovery rate, namely residual capacity ($C_{res}$), lost capacity ($C_l$), capacity after reconditioning ($C_{aft}$) and recovered capacity ($C_{rec}$). Here, we take the capacity value after cycling for $C_{res}$, which is slightly different from the capacity prior to reconditioning. In this case, $C_{rec}$ has accounted for capacity increase during the transition time between the end of cycling and reconditioning.

Furthermore, we also define two different parameters to quantify the capacity recovery rate, $r_1 = \frac{C_{rec}}{C_{res}} \times 100$ and $r_2 = \frac{C_{rec}}{C_l} \times 100$. $r_1$ would quantify the effect of residual capacity $C_{res}$ on the recoverable capacity. In contrast, $r_2$ quantifies the portion of lost capacity that is recovered. We used SHAP analysis[28] to quantify the importance of recovery parameters.

**Results and discussion**

Cycling and SoH assessment

The decay curves of discharge capacity with respect to cycle number at different currents are plotted in Fig. 2(a-l). The cycle life is defined as the cycle number where the cell capacity has reached 1.1 Ah at end of first life, which is around 73% SoH. There are clear correlations between the discharge capacity distribution and the cell cycling currents. We also found that the distribution of the battery cycle life with the same cycling current can be approximated by a Gaussian distribution, with symmetrical spreads (Fig. 2a-II). Given the discharge capacity, a higher current has a lower mean cycle life and a tighter spread. In contrast, a lower current has a higher mean cycle life and wider spread. For instance, the 3.0 A cycling current has a mean cycle life of about 100 cycles, with a range between 50 and 200 cycles. On the other extreme, the 1.0 A cycling current has a mean cycle life of around 900 cycles and a range between 600 to 1400 cycles.

Typical discharge curves and the corresponding relationships of differential capacity ($dQ/dV$) with respect to cell voltage at 2.0 A and 1.0 A are plotted in Fig. 2(b). A larger capacity drop between the 10$^{th}$ and 50$^{th}$ cycles is observed in the cell cycled at a higher current. There are distinctive features in the $dQ/dV$ curves in terms of peak positions and intensities at different currents and cycles. These features provide additional information about degradation compared to the features reported in high-power LFP cells.[12] In addition, Supplementary Information Fig. S4 presents the capacity decay and features of charge-discharge curves at different SoHs for a cell cycled at 1.0 A.

Table 1. RMSE and MAPE error metrics for linear and non-linear models that predict cycle life.

| Models | Mean of 5-fold CV RMSE (cycles) | | Mean of 5-fold CV MAPE (%) | |
|---|---|---|---|---|
| | Train-validation | Test | Train-validation | Test |
| LR model | 155 $\pm$ 7 | 173 $\pm$ 26 | 54.18 $\pm$ 2.29 | 65.07 $\pm$ 18.46 |
| GPR model | 109 $\pm$ 12 | 145 $\pm$ 48 | 23.77 $\pm$ 1.64 | 39.39 $\pm$ 15.80 |
| GBR model | 13 $\pm$ 1 | 140 $\pm$ 42 | 5.08 $\pm$ 0.45 | 16.84 $\pm$ 1.87 |

We used data-driven approaches to predict cycle life with both the linear model (LR) and the non-linear models (GPR and GBR). In line with Severson et al.'s study[12], we adopted RMSE and MAPE to assess model performance and generalizability. All the models take as a feature the log variance of the discharge curve difference between cycles $n_{start}$ and $n_{end}$, represented as $var(\Delta Q_{n_{start}-n_{end}}(U))$(Supplementary Information Fig. S5). Severson et al's study[12] has highlighted the predictive power of this variance, alongside the current and the capacity difference, ($\Delta Q_{n_{start}-n_{end}}$), between the two cycles. The cycle number set ($n_{start}$, $n_{end}$) is regarded as hyperparameters. While the dataset of 62 cells is relatively large in the academic battery community, it is still relatively small for model fitting and as a result, our data-driven models are prone to overfitting. To reduce overfitting, we implemented nested CV where model generalizability is evaluated in the outer loop while hyperparameter tuning is performed in the inner loop (Supplementary Information Fig. S3).

Table 1 shows the average RMSE and MAPE values for battery cycle life prediction obtained by evaluating the tuned models on the 5 train-validation/test sets followed by averaging the RMSE and MAPE values obtained; the standard deviation of each average RMSE or MAPE is also indicated in the table. The LR model reports an average test RMSE of 173 and an average test MAPE of 65.07% across the 5 test sets. These linear models generally perform more poorly than the nonlinear models (more details can be found in Supplementary Information Fig. S6). Besides linear models, we also explored nonlinear models such as GPR and GBR for cycle life forecasting. Both GPR and GBR models prove to be a better fit for our dataset than the LR models are. The average test RMSEs for the GPR and GBR stand at 145 and 140, respectively (Supplementary Information Figs. S7 and S8). In terms of test MAPE, the GBR performs the best with an average test MAPE as low as 16.84% across the 5 test sets.

The grid search for hyperparameters ($n_{start}$, $n_{end}$) in the inner loop using CV for all models is detailed in Supplementary Information Figs. S6 to S8. The GBR models display some variability in hyperparameters across the inner loop grid search. Fig. 2(c) presents an example of the parity plot of cycle life in which the GBR model is tuned and evaluated on the first fold of the outer loop (fold index = 0), which also happens to have the lowest RMSE and MAPE across all 5 outer loop folds. Fig. 2(d) provides an example of the contour plot of the validation set RMSE during grid search hyperparameter tuning; in this case, the optimal ($n_{start}$, $n_{end}$) was found to be (19, 77). Despite the variability in hyperparameters across different CV partitions, the GBR model's low test MAPE suggests that data-driven early prediction of the LFP cell's cycle life is feasible with our dataset, agreeing with the conclusion of Severson et al.'s prior study[12].

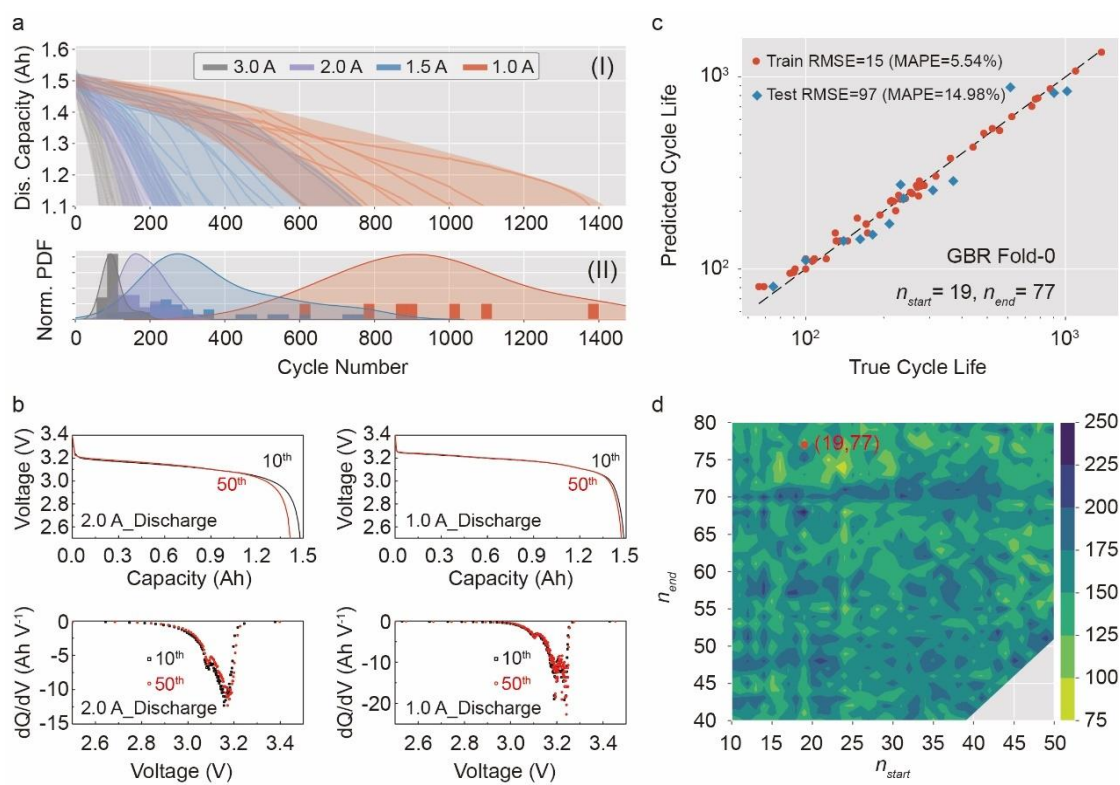

Fig. 2. (a-I) Discharge capacities of LFP/graphite 18650 cells at various cycling currents (3.0 A, 2.0 A, 1.5 A and 1.0 A). (a-II) Kernel density distribution plots of cycle numbers at a cut-off discharge capacity of 1.1 Ah (defined to be end of first life) for different cycling currents, (b) Typical discharge and differential capacity ($dQ/dV$) curves of the 10$^{th}$ and 50$^{th}$ cycles at 2.0 A and 1.0 A, (c) Example of parity plot of cycle life for a gradient boosting regression (GBR) model that is tuned and evaluated on the first fold of the outer loop, which also happens to give the lowest test RMSE and MAPE across all 5 outer loop folds. (d) Contour map of the validation set RMSE during grid search hyperparameter tuning, with the optimal hyperparameters ($n_{start}$, $n_{end}$) = (19, 77) marked by a red circle. Results from other splits are given in Supplementary Information Figs. S6 to S8.

Non-uniformity and aging

It is hardly possible to realize perfect uniformities during battery cell manufacturing. As illustrated in Fig. 3(a), either non-uniform materials distributions or uneven stacking of electrodes will lead to non-uniform ion transport and current density variations across the electrodes. Upon cycling, electrode materials may experience different volume changes due to the bias current/voltage. The side electrochemical reactions, e.g. SEI formation or Li plating, would be localized at different places on the electrodes as well. Compared with the intact anode and clean separator before cycling (Fig. 3b), the photos after cycling (Fig. 3c) show detached areas and white materials on the anode. Some dark strips of residual materials are also observed on the separator. It clearly demonstrates regionalized materials degradation and side reactions during the cell operation.

To have a closer examination, top view and cross-sectional view SEM images and elemental mappings of the anode before and after cycling are shown in Fig. 3(d) and 3(e). Before cycling, the elemental distributions of carbon (C) and fluorine (F) on the surface (Fig. 3d-i) are relatively uniform. The anode layers (~50 µm in thickness) are attached to the current collector well (Fig. 3d-ii). In contrast, the C and F distributions on the surface after cycling (Fig. 3e-I) indicate that certain F-rich materials form at some areas. The anode layer in Fig. 3(e-II) also becomes thicker (~58 µm) and its adhesion with the current collector becomes much worse. The observations of electrode materials at various magnifications confirm that the chemical and structural changes during cycling are uneven. Concurrently, the electric/electrochemical behaviors across the electrode will change differently. During cycling, some places will accumulate a higher potential for electric transport or electrochemical reaction as shown in Fig. 4(a-I) and 4(a-II). The electric potential changes are dynamic during battery operation. Although the high potential differences may decrease after resting (Fig. 4a-III), especially for high current operations, an electrical reconditioning step is recommended to further redistribute the Li ions, re-homogenize the electric potential differences and recover usable capacity (Fig. 4a-IV).

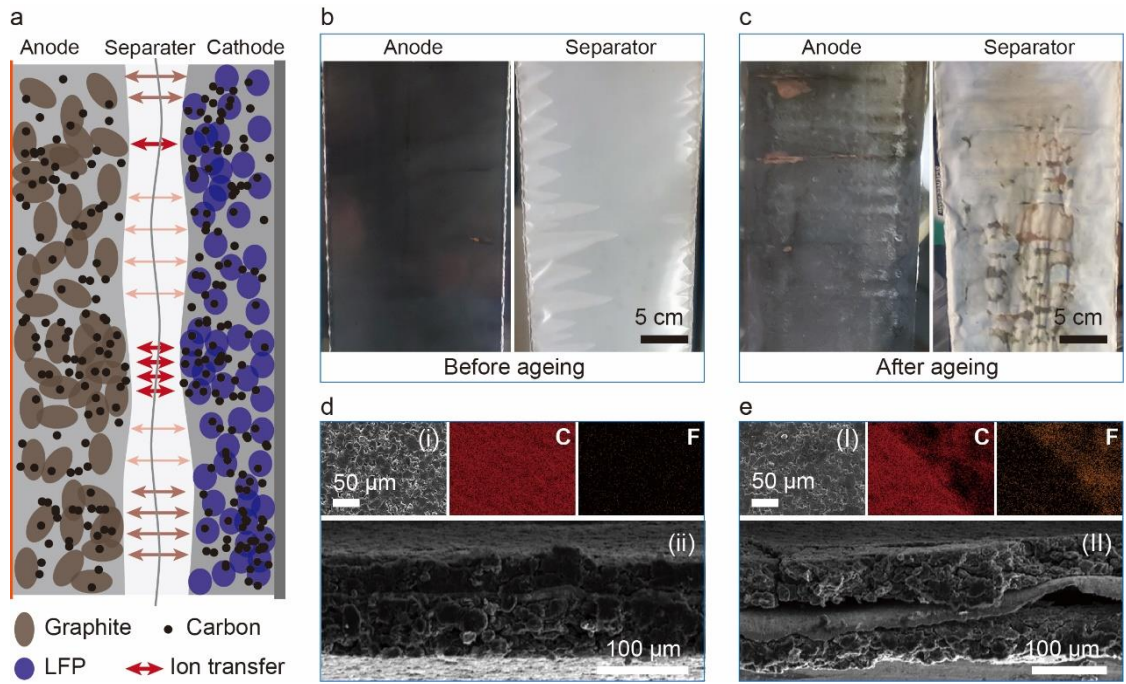

Fig. 3. (a) Illustration of non-uniformities of electrodes and variations of localized ion transfer in LFP/graphite cells. Optical images of graphite anode and separator (b) before and (c) after cycle aging. Top view SEM images (i and I) and elemental mappings of carbon (C) and fluorine (F), and cross-sectional view SEM images (ii and II) of graphite electrodes before (d) and after (e) cycle aging.

Using the ECM simulation (Fig. 4b), we can track the amount of lithium in each sub-cell during cycling (Fig. 4c), and reconditioning (Fig. 4d), with experiment-extracted model parameters presented in Supplementary Information Table S2. More details of the fitting process are provided in the 'ECM Fitting Procedure' section in Supplementary Information Tables S3-S5.

We can clearly see that the lithium in the two sub-cells diverge substantially from each other upon cycling and eventually settle to some steady level (Fig. 4c), with sub-cell 1 losing lithium, and sub-cell 2 gaining lithium. These divergences would illustrate the lithium inhomogeneity built up during cycling, by lithium movement across the bridge resistor $R_e$.

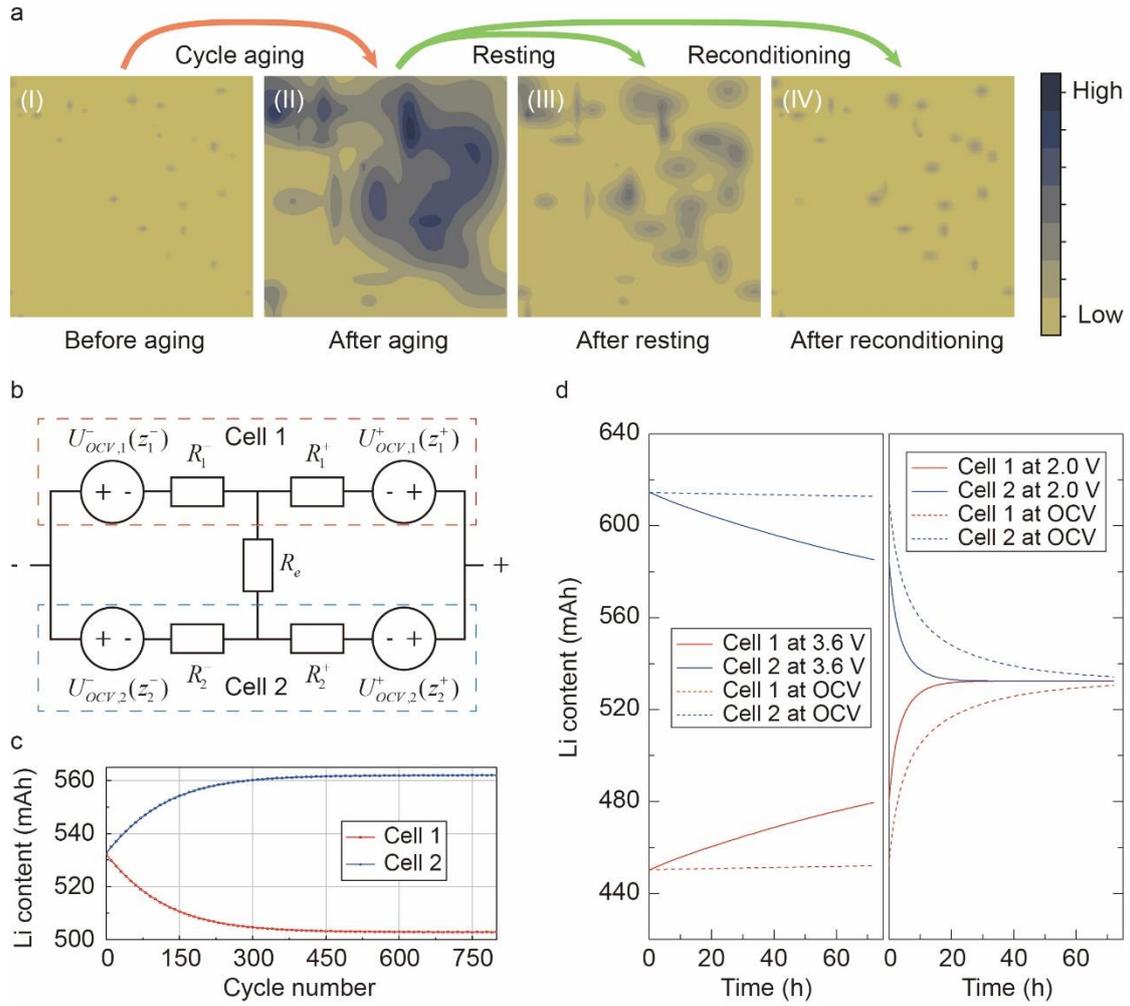

Fig. 4. (a) Schematic demonstration of non-uniform electric potential distribution before (I) and after (II) cycling and corresponding to resting (III) and reconditioning (IV), (b) Diagram of the equivalent circuit model (ECM) used, (c) Plots of time evolution of the amount of lithium in each sub-cell during cycling, (d) Lithium re-homogenization by holding the cell at constant voltage of 3.6 V (left) and 2.0 V (right). Re-homogenization by open circuit resting is shown by the dashed lines.

Capacity recuperation

Upon reconditioning, where the cell was held at a constant voltage (3.6 V or 2.0 V), the lithium ions in the two sub-cells are gradually equalized as shown in Fig. 4(d). These results suggest that the capacity recovery observed experimentally was contributed by the re-homogenization of Li ions. In addition, the capacity check of the cell at open-circuit resting was also conducted before and after the reconditioning in Supplementary Information Fig. S9.

There is a relaxation process during transit time, i.e after cycling and before constant voltage reconditioning. This relaxation recovers some of the cell's capacity, which saturates after a period. The constant voltage reconditioning brings back more capacity and the recovered capacity is stable during the following cycles as shown in Supplementary Information Fig. S9. This observation reflects the scenario of capacity recovery during open circuit resting and constant voltage reconditioning presented in Fig. 4(d).

Electric reconditioning treatments were carried out on the aged LFP cells as described in the experimental section. Various capacities together with recovery rates are defined and demonstrated in Fig. 5(a). The ECM can reasonably replicate the charge and discharge curves before and after reconditioning (Fig. 5b), with experimentally extracted parameters given in Supplementary Information Table S2. The fitting yields $R_e = 44\,\Omega$, which is two orders of magnitude larger than $R_1 = 0.15\,\Omega$ and $R_2 = 0.088\,\Omega$. Tracking the sub-cell lithium content with those parameters, the simulated Li inhomogeneity built-up during cycling is less than experimentally observed (Fig. 4c and 4d). This implies that there might be driving forces other than non-uniform resistance that promote the lateral lithium inhomogeneity, which the current ECM model had not considered yet. The ECM model does not incorporate any non-recoverable degradation mechanism which accounts for the substantial capacity fade from the nominal 1.5 Ah to about 1.0 Ah upon cycling. The electrode capacities obtained by parameter fitting also indicate significant loss of active materials and loss of lithium inventory, which may not be laterally uniform, and could also contribute to the observed lithium inhomogeneity. Despite these limitations, this simple ECM demonstrates the possibility of lithium non-uniformity build-up due to resistance difference and provides a mechanistic explanation for how inhomogeneous lithium can reduce the apparent capacity and how such capacity loss can be recovered by extreme voltage hold.

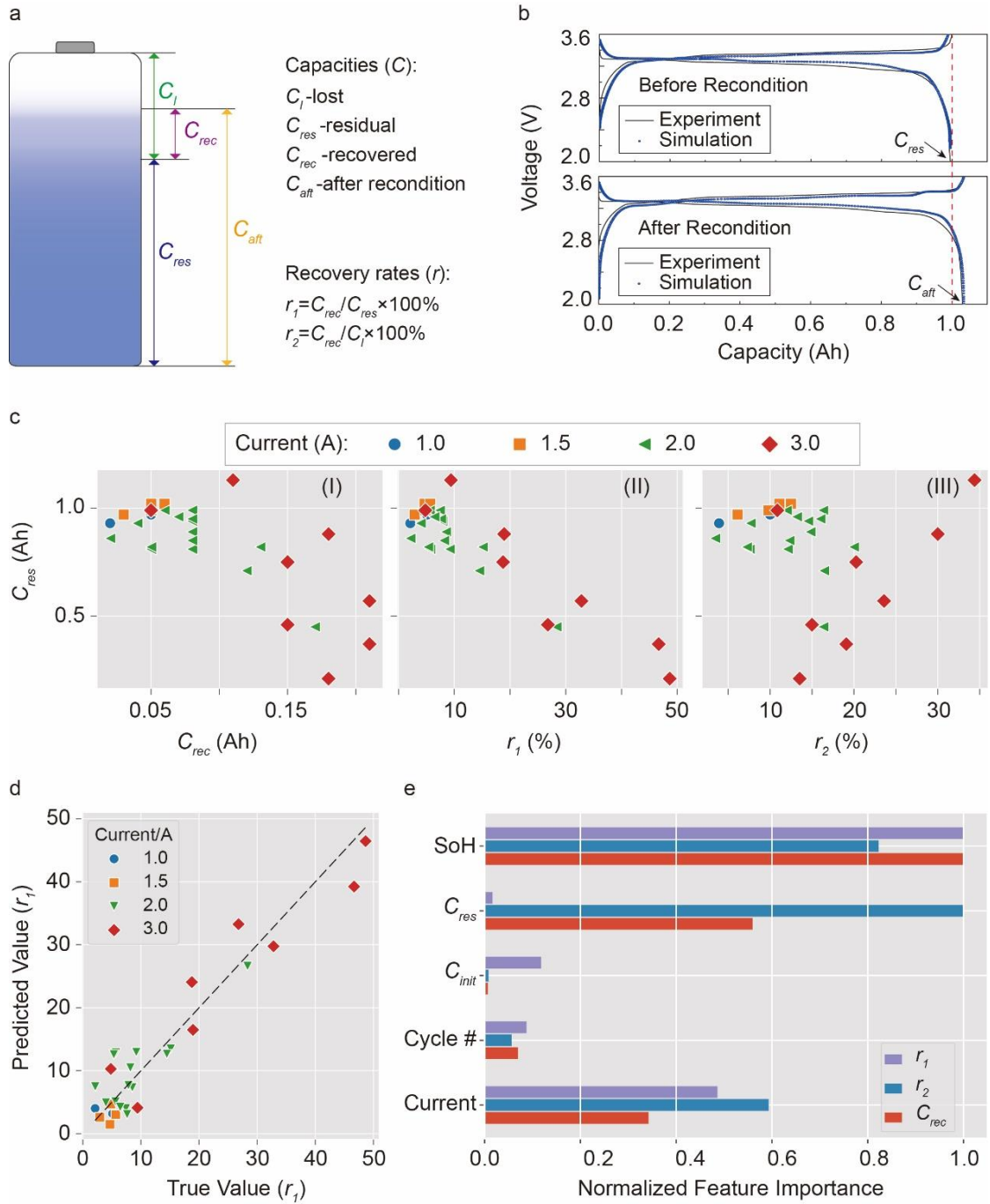

Fig. 5. (a) Illustration of various capacities and recovery rates during cell cycling and recuperation operation, (b) Experimental and simulated voltage profiles of capacity check before and after reconditioning, (c) Correlation of residual capacity ($C_{res}$) with respect to recovered capacity ($C_{rec}$) and recovery rates ($r_1$ and $r_2$), (d) Parity plot of observed and predicted recovery rates $r_1$ using a linear regression model, and (e) Bar graph showing the importance of different parameters (SoH, $C_{res}$, $C_{init}$, cycle number (Cycle #) and Current) for predicting the capacity recovery effects.

Reconditioning recovered between 0.02 Ah to 0.21 Ah, which translates to 3-34% recovery of the loss capacity (Fig. 5c-I and -II). There is a negative correlation between the residual capacity $C_{res}$ and the recovered capacity $C_{rec}$ (Fig. 5c-I), as higher cycling rates seems to recover more (i.e 2.0 A and 3.0 A) than those that cycled with lower rates (i.e 1.0 A and 1.5 A). In general, the recovery metrics ($C_{rec}$, $r_1$) increase with loss in capacity (lower residual capacity $C_{res}$) (Fig. 5c). This is more prominent with $r_1$, which has a negative linear correlation with $C_{res}$. On the other hand, the relationship between $C_{res}$ and $r_2$ is complex. There is a negative correlation for cells that are cycled with 1.0-2.0 A current, but a positive correlation for cells that are cycled with 3.0 A current. Cycling with 3.0 A (~2C) current can be considered harsh, so we would expect it to decrease the recoverable capacity. However, some cells from the high cycling rate showed relatively high $r_2$ recovery rates (Fig. 5c-III).

The results for other parameters, like the initial capacity $C_{init} = C_{res} + C_l$, SoH and cycle number (#), are provided in Supplementary Information Fig. S10.

Overall cell capacity recoverability can be well predicted from the cycling parameters (current and number of cycles) and states of the cells ($C_{init}$, SoH, $C_{res}$). Here, a linear regression model performs very well in predicting the cell's recoverability in terms of the recovery rate $r_1$ (Fig. 5d). Prediction of other recovery indicators $C_{rec}$, and $r_2$ also works well with a linear regression model (Supplementary Information Fig. S11). In Fig. 5(e), SHAP[28] analysis showed that SoH is the most important parameter to affect $C_{rec}$, $r_1$ and $r_2$. More details of their correlations are given in Supplementary Information Fig. S12. In the case where the history of cycling is unknown, $r_1$ seems to be the best metric to estimate the cell's capacity recoverability. $r_1$ correlates well linearly with SoH (Supplementary Information Fig. S10) and is reasonably less sensitive to cycling rates.

**Conclusion**

Data-driven health estimation and statistical analysis of capacity recovery of aged Li-ion batteries play significant roles in their second-life applications. Machine learning modeling on our cycling dataset of 62 LFP/graphite 18650 cells showed that it is possible to make early prediction of cell cycle life. A gradient boosting regression model is used to demonstrate the feasibility of early prediction of battery cycle life with a test RMSE of 140 ± 42 cycles and test MAPE of 16.84% ± 1.87%, using data from the first 80 cycles. Non-uniformities are detected in battery cells, which are proposed as a reason for performance degradation and capacity loss. The equivalent circuit model simulation suggests lateral lithium inhomogeneity can build up upon cycling due to uneven internal resistance, while such inhomogeneity contributes to apparent capacity loss. The simulation also indicates that extreme voltage hold can be effective in re-homogenizing lithium and recovering the portion of lost capacity caused by lithium inhomogeneity. Recuperation effectiveness is validated by reconditioning experiments with considerable capacity recovery rates at different conditions. Further data analysis reveals the importance of SoH in determining the capacity recovery of aged LFP batteries. This study demonstrates the promising applications of data-driven methods and equivalent circuit modeling in battery diagnostics and highlights their significance in the emerging field of second-life batteries.

**Acknowledgement**

This research is supported by A*STAR (project no. C210812038) and EMA-EP011-SLEP-001 (SC25/21-708712). R.I.M. acknowledges funding from Accelerated Materials Development for Manufacturing Program at A*STAR via the AME Programmatic Fund by the Agency for Science, Technology and Research under Grant (Project Code A188b0043). J.L., Y.Z. and E.K. acknowledge funding by Agency for Science, Technology and Research (A*STAR) under the Career Development Fund (C210112037). The authors acknowledge Kedar Hippalgaonkar's helpful discussion during project initiation.

Supplementary information for

# Health diagnosis and recuperation of aged Li-ion batteries with data analytics and equivalent circuit modeling

Riko I Made[a], Jing Lin[b], Jintao Zhang[a], Yu Zhang[b], Lionel C. H. Moh[a], Zhaolin Liu[a], Ning Ding[a], Sing Yang Chiam[a], Edwin Khoo[b,*], Xuesong Yin[a,*] and Guangyuan Wesley Zheng[c]*

[a] Institute of Materials Research and Engineering (IMRE), Agency for Science, Technology and Research (A*STAR), 2 Fusionopolis Way, Innovis #08-03, Singapore 138634, Republic of Singapore.
Email: yinxs@imre.a-star.edu.sg
[b] Institute for Infocomm Research (I²R), Agency for Science, Technology and Research (A*STAR), 1 Fusionopolis Way, #21-01 Connexis, Singapore 138632, Republic of Singapore.
Email: edwin_khoo@i2r.a-star.edu.sg
[c] Posh Robotics, 3501 Breakwater Court, Hayward, CA 94545, United States.
Email: wesley@poshrobotics.com

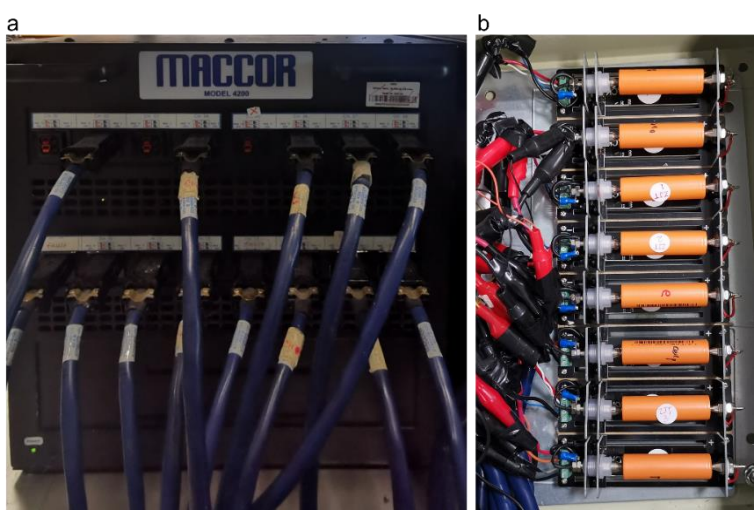

Fig. S1. (a) Battery tester and (b) cell holder used for cell cycling.

Table S1. Cell cycling information and capacity recovery after reconditioning treatments. $C_i$, $C_{res}$, $C_{aft}$ and $C_{rec}$ represent initial capacity, residual capacity, capacity after reconditioning and recovered capacity, respectively. $r_1$ and $r_2$ are recovery rates defined as $r_1 = \frac{C_{rec}}{C_{res}} \times 100\%$ and $r_2 = \frac{C_{rec}}{C_i} \times 100\%$.

| Cell ID | Current/A | Cycle # | $(C_i+C_{res})$/Ah | $C_{res}$/Ah | $C_{aft}$/Ah | $C_{rec}$/Ah | $r_1$/% | $r_2$/% |
|---|---|---|---|---|---|---|---|---|
| 72 | 1.0 | 918 | 1.47 | 0.97 | 1.03 | 0.05 | 5.10 | 10.00 |
| 73 | 1.0 | 1368 | 1.44 | 0.93 | 0.97 | 0.02 | 2.11 | 3.92 |
| Cella | 1.5 | 600 | 1.46 | 0.97 | 1.06 | 0.03 | 2.91 | 6.12 |
| CellG | 1.5 | 506 | 1.50 | 0.99 | 1.09 | 0.05 | 4.81 | 9.80 |
| CellJ | 1.5 | 600 | 1.47 | 1.02 | 1.12 | 0.05 | 4.67 | 11.11 |
| CellT | 1.5 | 289 | 1.50 | 1.02 | 1.12 | 0.06 | 5.66 | 12.50 |
| Cellf | 2.0 | 200 | 1.49 | 0.99 | 1.13 | 0.06 | 5.61 | 12.00 |
| Cellg2 | 2.0 | 369 | 1.43 | 0.89 | 1.02 | 0.08 | 8.51 | 14.81 |
| Cellh2 | 2.0 | 230 | 1.48 | 0.99 | 1.16 | 0.08 | 7.41 | 16.33 |
| Cellj2 | 2.0 | 292 | 1.49 | 0.94 | 1.10 | 0.08 | 7.84 | 14.55 |
| Cellk2 | 2.0 | 268 | 1.47 | 0.82 | 0.99 | 0.13 | 15.12 | 20.00 |
| Celll2 | 2.0 | 200 | 1.44 | 0.71 | 0.95 | 0.12 | 14.46 | 16.44 |
| Cellm | 2.0 | 200 | 1.45 | 0.95 | 1.13 | 0.08 | 7.62 | 16.00 |
| Cello | 2.0 | 245 | 1.47 | 0.81 | 0.95 | 0.08 | 9.20 | 12.12 |
| Cellp2 | 2.0 | 368 | 1.50 | 0.85 | 1.06 | 0.08 | 8.16 | 12.31 |
| Cellq2 | 2.0 | 252 | 1.49 | 0.45 | 0.77 | 0.17 | 28.33 | 16.35 |
| Cellr2 | 2.0 | 349 | 1.47 | 0.81 | 0.93 | 0.05 | 5.68 | 7.58 |
| Cells2 | 2.0 | 255 | 1.51 | 0.82 | 0.99 | 0.05 | 5.32 | 7.25 |
| Cellu2 | 2.0 | 200 | 1.44 | 0.86 | 0.96 | 0.02 | 2.13 | 3.45 |
| Cell81 | 2.0 | 200 | 1.49 | 0.96 | 1.16 | 0.07 | 6.42 | 13.21 |
| Cell82 | 2.0 | 200 | 1.45 | 0.93 | 1.05 | 0.04 | 3.96 | 7.69 |
| Cell15 | 3.0 | 100 | 1.48 | 0.88 | 1.13 | 0.18 | 18.95 | 30.00 |
| Cell16 | 3.0 | 100 | 1.45 | 1.13 | 1.28 | 0.11 | 9.40 | 34.38 |
| CellC10 | 3.0 | 240 | 1.47 | 0.37 | 0.66 | 0.21 | 46.67 | 19.09 |
| CellC11 | 3.0 | 132 | 1.45 | 0.99 | 1.09 | 0.05 | 4.81 | 10.87 |
| CellC12 | 3.0 | 200 | 1.46 | 0.57 | 0.85 | 0.21 | 32.81 | 23.60 |
| CellC14 | 3.0 | 200 | 1.54 | 0.21 | 0.55 | 0.18 | 48.65 | 13.53 |
| CellC16 | 3.0 | 140 | 1.46 | 0.46 | 0.71 | 0.15 | 26.79 | 15.00 |
| CellC7 | 3.0 | 130 | 1.49 | 0.75 | 0.95 | 0.15 | 18.75 | 20.27 |

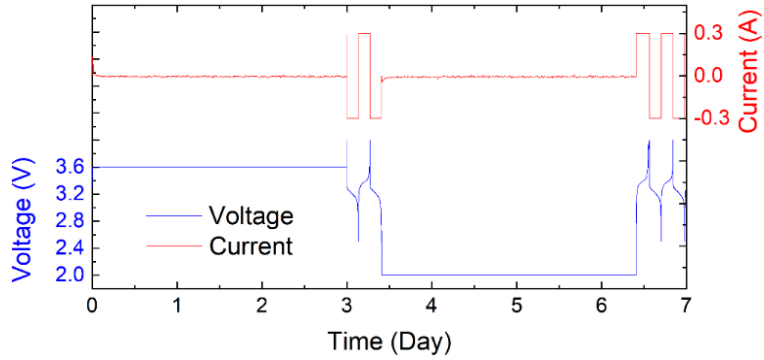

Fig. S2. Typical current and voltage profiles during reconditioning and capacity check.

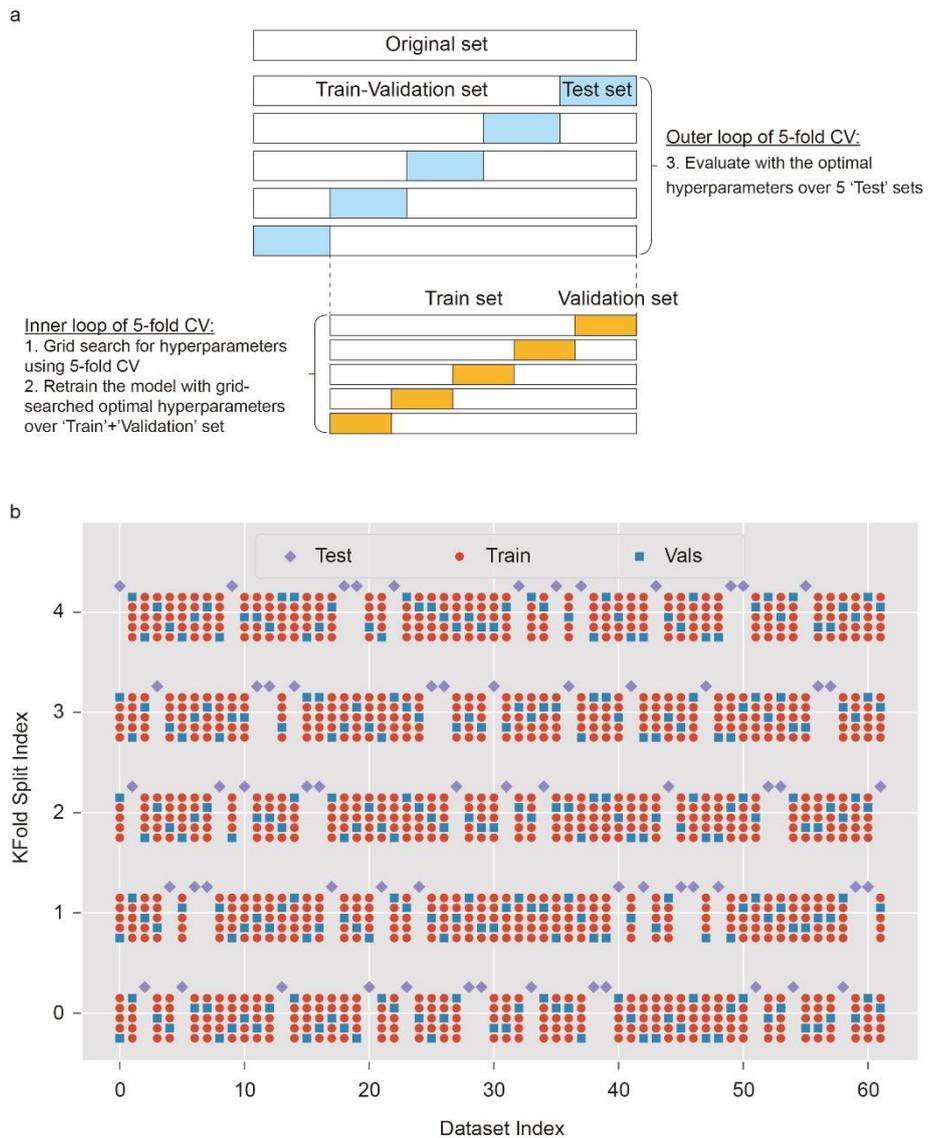

Fig. S3. (a) Dataset splitting scheme between outer loop (Train-Validation, Test) and inner loop (Train and Validation set) for nested cross-validation (CV). The inner loop is used to find the optimal hyperparameters across the 5-fold Train–Validation set. The model with the best performance is retained and finally retrained on the Train-Validation set, which is then evaluated on the respective

Test set. (b) The generated data partition of the Train, Validation and Test set, is represented with red-circle, blue-square, and purple-diamond symbols, respectively. The dataset index represents the row numbers of the dataset.

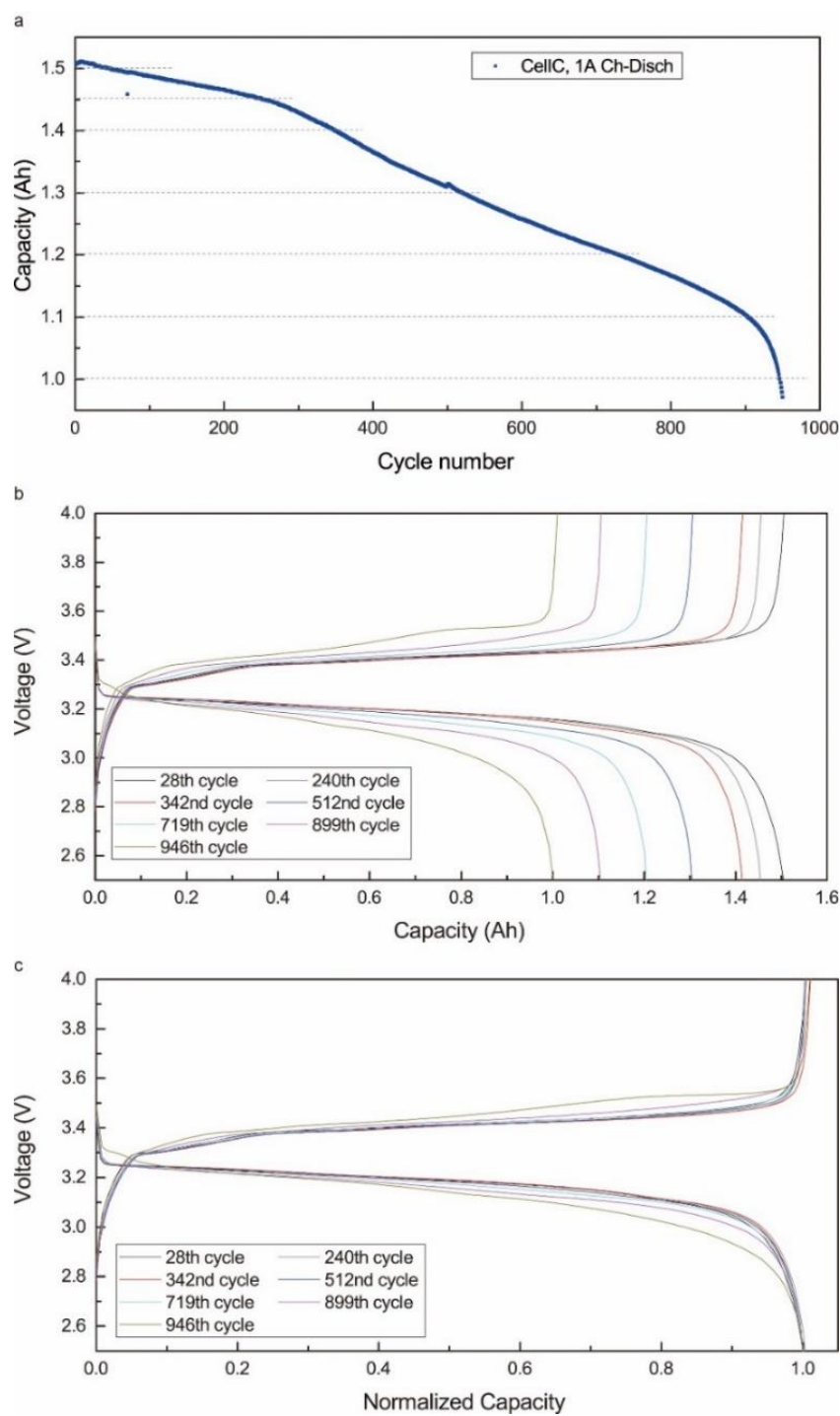

Fig. S4. Cycling behaviors of Cell C cycled at 1.0 A. (a) Capacity decay with respect to cycle number. (b) Charge-discharge curves at different capacities of 1.50 Ah ($28^{th}$ cycle), 1.45 Ah ($240^{th}$ cycle), 1.40 Ah ($342^{nd}$ cycle), 1.30 Ah ($512^{th}$ cycle), 1.20 Ah ($719^{th}$ cycle), 1.10 Ah ($899^{th}$ cycle) and 1.00 Ah ($946^{th}$ cycle). (c) Normalized charge-discharge profiles of Cell C at different capacities in (b).

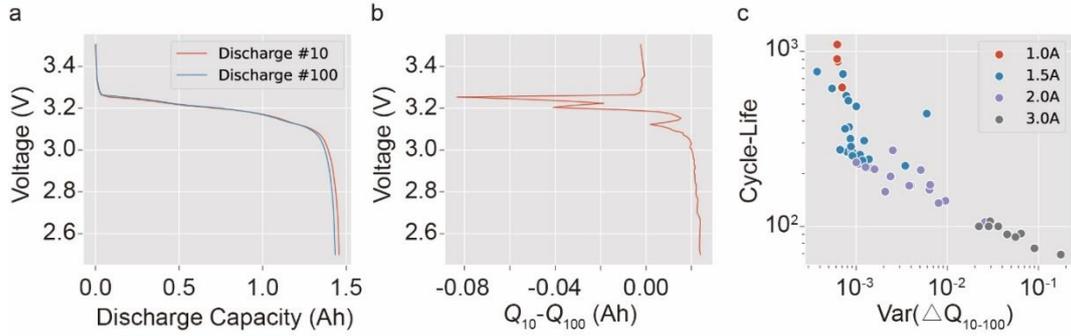

Fig. S5. Features $var(\Delta Q_{n_{start}-n_{end}})$. (a) Discharge voltage curves from two selected cycles; here for illustration, we plot those from cycles 10 and 100. (b) Calculated capacity differences from the two discharge cycles $\Delta Q$. (c) Cell cycle life as a function of variance of $\Delta Q$ ($var(\Delta Q)$), which shows an exponential decay relation.

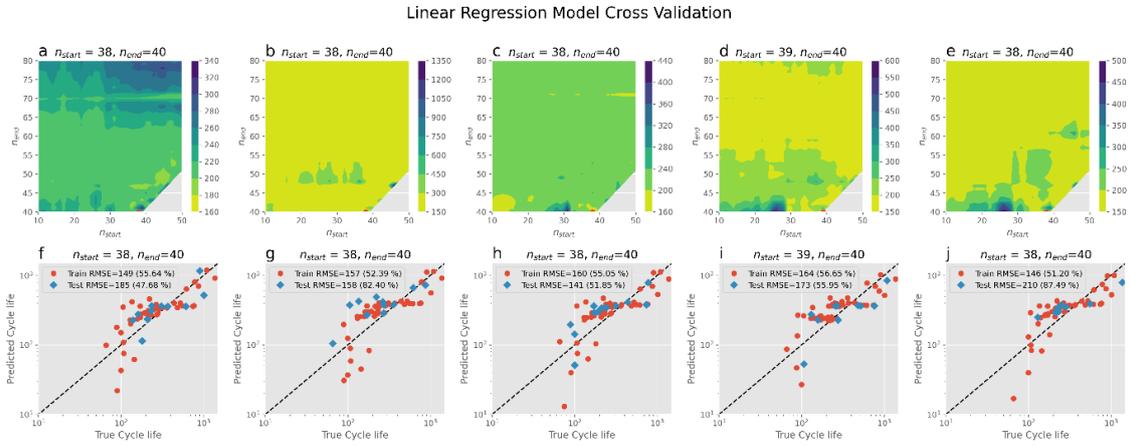

Fig. S6. (a-e) Linear regression contour map of the validation set's RMSE from 5-fold train-validation, with the red circular dot representing the best RMSE for each fold, respectively. (f-j) Parity plot of the model performance, with the MAPE given inside the brackets in the legend.

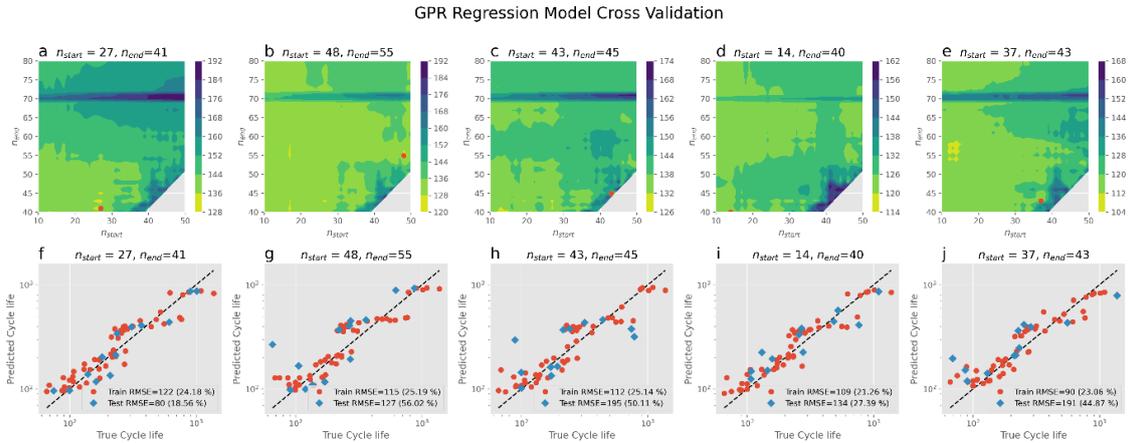

Fig. S7. (a-e) GPR contour map of the validation set's RMSE from 5-fold train-validation, with the red circular dot representing the best RMSE for each fold, respectively. (f-j) Parity plot of the model performance, with the MAPE given inside the brackets in the legend.

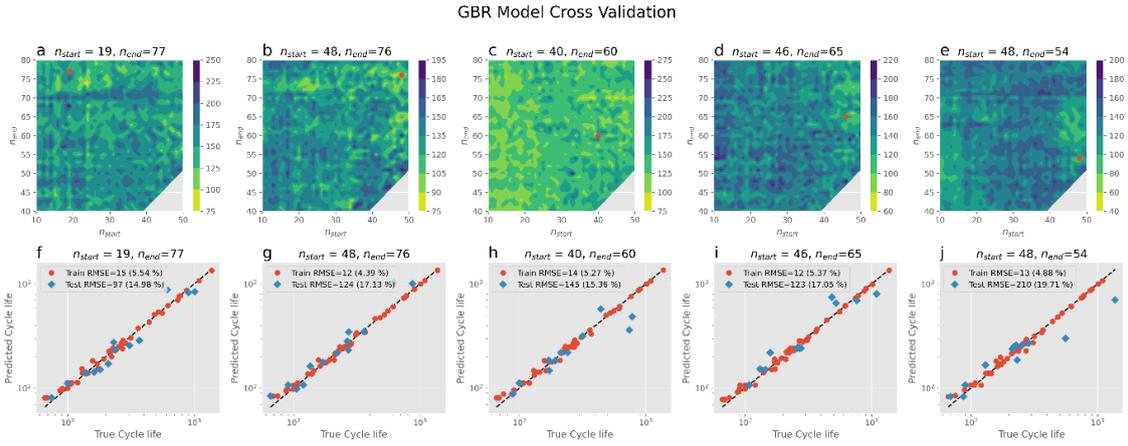

Fig. S8. (a-e) GBR contour map of the validation set's RMSE from 5-fold train-validation, with the red circular dot representing the best RMSE for each fold, respectively. (f-j) Parity plot of the model performance, with the MAPE given inside the brackets in the legend.

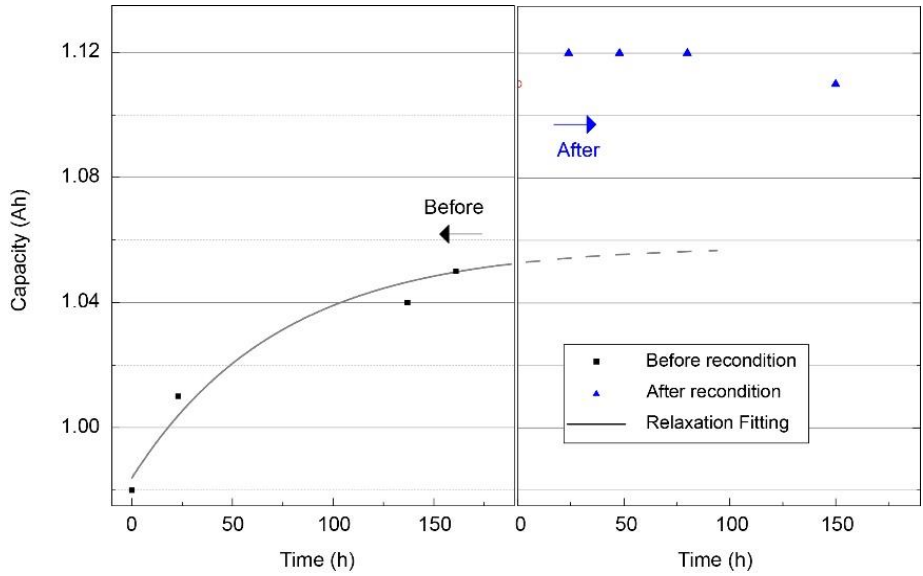

Fig. S9. Capacity changes of a cycled cell before and after reconditioning during open circuit voltage (OCV) storage. The fitted curve of rest before reconditioning follows y = 1.058(1-0.07e$^{-x/75.702}$).

Table S2. Parameters for the used ECM.

| Parameter | Value |
| --- | --- |
| $R_1^{\pm}$ | 149.85 mΩ |
| $R_2^{\pm}$ | 88.15 mΩ |
| $R_e$ | 44514.26 mΩ |
| $R_0 = \left(2k/(k+1)\right)R_2^{\pm}$ | 111 mΩ |
| $k = R_1^{\pm}/R_2^{\pm}$ | 1.7 |
| $k_e = R_e/R_2^{\pm}$ | 505 |
| $U^-_{OCP,1/2}(z^-_{1/2})$ | Graphite (MCMB) Open Circuit Potential from Plett 2015.[1] 2.4V to 4.6V for $z^-$ from 1 to 0. |
| $U^+_{OCP,1/2}(z^+_{1/2})$ | LFP Open Circuit Potential from Plett 2015.[1] 0.0V to 3.0V for $z^+$ from 1 to 0. |
| $Q^-_{max}$ | 1.227Ah |
| $r_{N/P} = Q^-_{max}/Q^+_{max}$ | 1.064 |
| $z^-_1(t_0)$ | 0.001 |
| $z^+_1(t_0)$ | 0.752 |
| $z^-_2(t_0)$ | 0.104 |
| $z^+_2(t_0)$ | 0.983 |

Table S3. Cell cycling and reconditioning protocols. The number of cycles 919 here refers to a particular cell cycled at 1A whose data are fit by the ECM used.

| Operation | Protocol | #Rep |
|---|---|---|
| Cycling | 1. 1A CC charge to 4V -> 2min rest<br>1. 1A CC discharge to 2.5V -> 2min rest | 919 |
| Checkup | 1. 0.3A CC charge to 3.6V -> 2min rest<br>2. 0.3A CC discharge to 2V - 2min rest | 2 |
| Voltage hold | 2A CC charge to 3.6V -> 3.6V hold for 3 days -> 10min rest | 1 |
| Voltage hold | 2A CC discharge to 2V -> 2V hold for 3 days -> 10min rest | 1 |
| Checkup | 3. 0.3A CC charge to 3.6V -> 10min rest<br>4. 0.3A CC discharge to 2V - 10min rest | 2 |

Table S4. The measured and simulated discharge capacity at the last complete cycle (#918) during cycling, the second cycle of the pre-reconditioning capacity checkup, and the second cycle of the post-reconditioning checkup.

| | Cycle 918 | Pre checkup | Post checkup |
|---|---|---|---|
| **Experiment dis Q [Ah]** | 0.967 | 0.995 | 1.030 |
| **Simulation dis Q [Ah]** | 0.972 | 0.995 | 1.034 |

**ECM Fitting Procedure**

The ECM of Fig. 4(b) has nine unknown parameters as listed in Table S5, and with each set of parameter values, it can be time-integrated by scipy.integrate.solve_ivp in the SciPy library to obtain the voltage response to any imposed voltage or current protocol. To fit the parameters, we use the protocols listed in Table S3 chained in a sequence, except that we only simulate the last three cycles (cycle 917 to 919) of the cycling part to reduce simulation costs. The initial time $t_0$ is set to the beginning of these initial three cycles. The model voltage responses for cycle 918, the second pre-reconditioning checkup cycle, and the two post-reconditioning checkup cycles are compared to their experimental counterparts and the optimal parameters are to minimize the mean squared errors based on these voltage data. This nonlinear least-square problem is then solved by the global optimization algorithm of differential evolution implemented as scipy.optimize.differential_evolution in the SciPy library, with the parameters constrained to the ranges listed in Table S5 chosen according to prior physical knowledge of the cell and some preliminary trial-and-error parameter exploration. The obtained parameter values are listed in Table S2.

Table S5. ECM parameters to be fit to experiment data with their prior ranges.

| Parameter | Range |
|---|---|
| $R_2^{\pm}$ | [10mΩ, 300mΩ] |
| $k = R_1^{\pm}/R_2^{\pm}$ | [1, 8] |
| $k_e = R_e/R_2^{\pm}$ | [500, 2000] |
| $Q_{max}^{-}$ | [1.1Ah, 2.2Ah] |
| $r_{N/P} = Q_{max}^{-}/Q_{max}^{+}$ | [1, 2] |
| $z_1^{-}(t_0)$ | [1e-8, 0.3] |
| $z_1^{+}(t_0)$ | [0.7, 1 - 1e-8] |
| $z_2^{-}(t_0)$ | [1e-8, 0.3] |
| $z_2^{+}(t_0)$ | [0.7, 1 - 1e-8] |

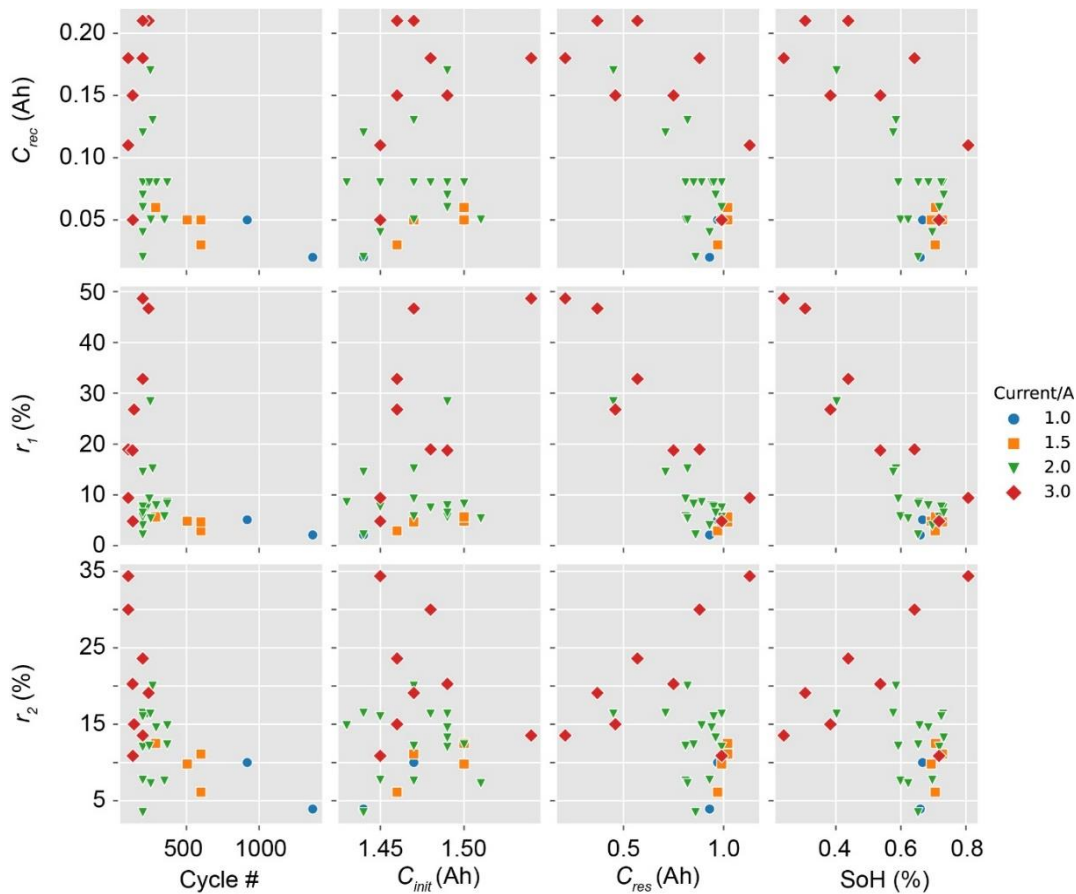

Fig. S10. Correlation between defined recovery metrics $r_1$, $r_2$ and $C_{rec}$, with the cell's history and end of cycle state, i.e. cycle number (Cycle #), initial capacity ($C_{init}$), residual capacity ($C_{res}$), and state of health (SoH). Additionally, we use different symbols with different colors to indicate various cycling currents.

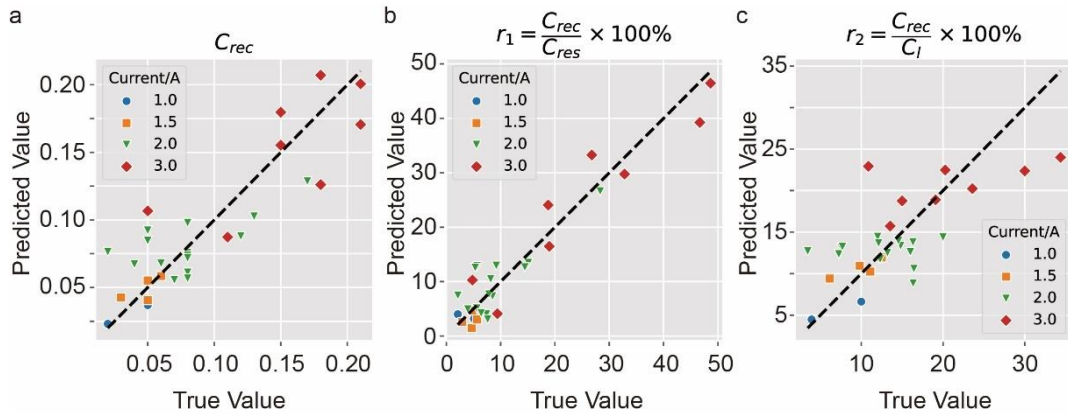

Fig. S11. Linear regression modeling of recovery rates with respect to SoH, residual capacity $C_{res}$, initial capacity $C_{init}$, cycle number, and cycling current. (a) Recovered capacity ($C_{rec}$), (b) recovery rate $r_1$, (c) recovery rate $r_2$.

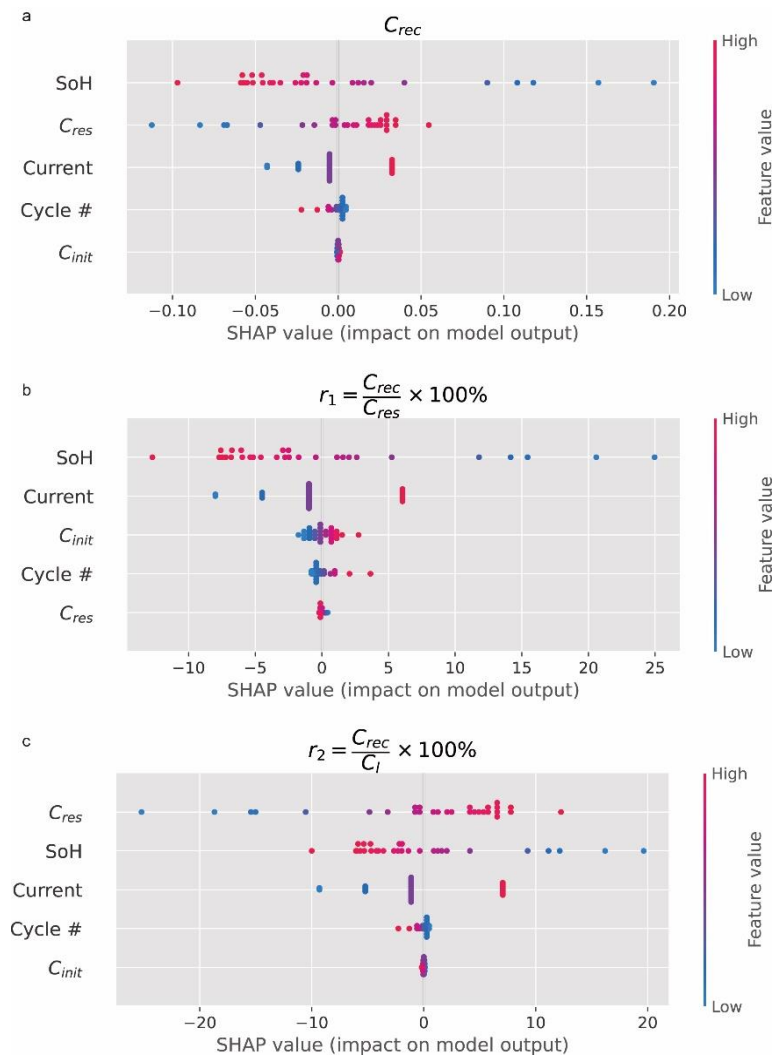

Fig. S12. SHAP analysis of the recovery metrics (a) $C_{rec}$, (b) $r_1$ and (c) $r_2$ with respect to cell's history and end of cycle state, i.e. cycle number (Cycle #), initial capacity ($C_{init}$), residual capacity ($C_{res}$), state of health (SoH) and current (Current).